\documentclass[preprint,noshowpacs,preprintnumbers,amsmath,amssymb,prl,superscriptaddress]{revtex4}

\usepackage{graphicx}
\usepackage{dcolumn}
\usepackage{bm}
\usepackage{epsfig}
\usepackage{natbib}
\usepackage{color}
\usepackage{xcolor}

\newcommand{\rs}{r_\text{s}} 

\def  \bsig    {\mbox{\boldmath$\sigma$}}

\begin{document}
\title{Interplay of  spin-orbit coupling and Coulomb interaction in ZnO-based electron system}
\author{D.~Maryenko}
\email{maryenko@riken.jp}
\affiliation{RIKEN Center for Emergent Matter Science(CEMS), Wako 351-0198, Japan}
\author{M.~Kawamura}
\affiliation{RIKEN Center for Emergent Matter Science(CEMS), Wako 351-0198, Japan}
\author{A.~Ernst}
\affiliation{Institute for Theoretical Physics, Johannes Kepler University, 4040 Linz, Austria}
\affiliation{Max Planck Institute of Microstructure Physics, D-06120 Halle, Germany}
\author{V.~K.~Dugaev}
\affiliation{Department of Physics and Medical Engineering, Rzesz\'{o}w University of Technology, 35-959 Rzesz\'{o}w, Poland}
\author{E.~Ya.~Sherman}
\affiliation{Department of Physical Chemistry, University of the Basque Country UPV/EHU, Apartado 644, Bilbao 48080,  Spain}
\affiliation{Ikerbasque, Basque Foundation for Science, Bilbao, Spain}
\author{M.~Kriener}
\affiliation{RIKEN Center for Emergent Matter Science(CEMS), Wako 351-0198, Japan}
\author{M.~S.~Bahramy}
\altaffiliation{Current address: Department of Physics and Astronomy, The University of Manchester, Oxford Road, Manchester M13 9PL, United Kingdom}
\affiliation{Department of Applied Physics and Quantum-Phase Electronics Center (QPEC), The University of Tokyo, Tokyo 113-8656, Japan}

\author{Y.~Kozuka}
\affiliation{Research Center for Magnetic and Spintronic Materials, National Institute for Materials Science (NIMS), Tsukuba 305-0047, Japan}
\affiliation{JST, PRESTO, Kawaguchi, Saitama 332-0012, Japan}

\author{M.~Kawasaki}
\affiliation{RIKEN Center for Emergent Matter Science(CEMS), Wako 351-0198, Japan}
\affiliation{Department of Applied Physics and Quantum-Phase Electronics Center (QPEC), The University of Tokyo, Tokyo 113-8656, Japan}

\maketitle

\textbf{Spin-orbit coupling (SOC) is pivotal for various fundamental spin-dependent phenomena in solids and their technological applications. In semiconductors, these phenomena have been so far studied  in relatively weak electron-electron interaction regimes, where the single electron picture holds. However, SOC can profoundly compete against Coulomb interaction, which could lead to the emergence of unconventional electronic phases. Since SOC depends on the electric field in the crystal including contributions of itinerant electrons, electron-electron interactions can modify this coupling.  Here we demonstrate the emergence of SOC effect in a high-mobility two-dimensional electron system in a simple band structure MgZnO/ZnO semiconductor. This electron system features also strong electron-electron interaction effects. By changing the carrier density with Mg-content, we tune the SOC strength and achieve its interplay with electron-electron interaction. These systems pave a way to emergent spintronic phenomena in strong electron correlation regime and to the formation of novel quasiparticles with the electron spin strongly coupled to the density. }

Spin-orbit coupling is a single particle relativistic effect producing in atomic physics a bilinear interaction between the electron spin  and its orbital momentum. 
In solids the SOC is transformed into a symmetry-permitted coupling between the orientation of the electron spin and its crystal momentum. 
This coupling can lead to spin-momentum locking and establishes a spin-dependent band structure influenced by the crystal symmetry. 
Prominent examples here are  the Rashba and Dresselhaus couplings, whose appearance requires the breaking of the structural and crystal inversion symmetries. 
By contrast, Coulomb interaction dictates collective electron  behaviour in solids, e.g., by establishing a Fermi liquid or a Mott insulator, and can also generate spin-polarized phases due to the Stoner instability~\cite{StonerFerromagnetism}. 
Thus, SOC orients electron spin with respect to its momentum while the Coulomb interaction can counteract  by aligning the spins in one direction, e.g., by producing a spin-depended exchange interaction.  
The usual single particle description of SOC-related effects in the presence of Coulomb interaction is poorly applicable, since the relativistic effect  on quasiparticle excitations in strongly interacting systems is not known. 
Yet, the interplay of two mechanisms for spin orientation is suggested to have diverse manifestations encompassing the emergence of topological phases, spin textures, etc.~\cite{SpinTexturePrediction2017, Ashrafi2012, Witczak-Krempa}. 
An experimental realisation of a system that shows both strong interaction between electrons, e.g., in the form of a Fermi liquid, and spin-orbit coupling is challenging. 
It requires a system with sufficiently strong relativistic effects to unfold the role of spin-orbit coupling and with a high mobility at a low carrier density to enhance the Coulomb interaction effect.

Here we demonstrate a realisation of such a regime in the two-dimensional electron system (2DES) at the Mg$_x$Zn$_{1-x}$O/ZnO interface.  The SOC effect is identified from the beatings of the Shubnikov-de Haas oscillations (SdH) in conductivity, which varies with the electron density $N$. Upon lowering $N$ the system shows an enhancement of the electron effective mass, attributed to electron-electron interaction.  Thus, we can tune the interplay between two interaction mechanisms and achieve an unprecedented interaction regime for 2DESs, where the emergence of novel quantum states is anticipated.

\begin{figure}[!htb]
\includegraphics{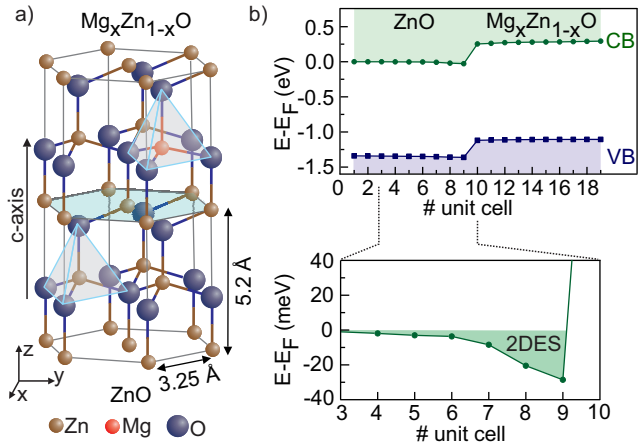}
\caption{\label{Fig1} \textbf{Electronic structure of Mg$_x$Zn$_{1-x}$O/ZnO interface}. \textbf{a)} Schematic view of high mobility MgZnO/ZnO heterostructure. Both wurtzite crystal structure 
of ZnO and  Mg$_x$Zn$_{1-x}$O/ZnO interface breaks the inversion symmetry. \textbf{b)}  The interface band structure is calculated using self consistent Green function method for semi-infinite systems considering x=5$\%$, a typical Mg content in the heterostructures. The conduction band (CB) of ZnO lowers at the interface forming the confinement potential for high mobility electrons.  The size of the band gap in ZnO and  Mg$_x$Zn$_{1-x}$O is underestimated due to the lack of the conventional density functional theory.}
\end{figure}

We start with the discussion of the 2DES formation, since it is central for  tuning the interplay between two interaction mechanisms. The 2DES  is realised in the $c$-plane of wurtzite ZnO by interfacing it with Mg$_x$Zn$_{1-x}$O (Fig.~\ref{Fig1}(a)). Its formation is validated by our first-principles calculations, modelling  the interface between two semi-infinite systems, ZnO and Mg$_{x}$Zn$_{1-x}$O (Fig.~\ref{Fig1}(b)). While Mg substitutes Zn stoichiometrically, its position is  shifted from the original Zn atom position resulting in $c$-axis shrinking of the Mg$_x$Zn$_{1-x}$O layer. This and the different chemical environment brought in by Mg atoms lead to a polarization discontinuity at the  Mg$_x$Zn$_{1-x}$O/ZnO interface, which  is compensated by accumulating electrons at the interface. Respectively, the electron density depends on the Mg-content \cite{MgCallibration}.

In such a wurtzite heterostructure the electrons are allowed to be polarised by the spin-orbit interaction, since both structural and crystal inversion symmetries are broken. The corresponding Hamiltonian for electrons in the $c$-plane is:
\begin{equation}
H_{\rm SOC}=\left[ \alpha_R+\gamma(b\langle k_{z}^{2}\rangle -k_{\parallel}^2)\right](\sigma_{x}k_{y}-\sigma_{y}k_{x}),
\end{equation}
where $\alpha_R$ and $\gamma$ are the Rashba and Dresselhaus coefficients respectively \cite{SOIWurtzite1, Cardona1996, Chuang1996}. Here $k_{z}=-i\partial/\partial z$ acting on the electron wavefunction with $\langle\ldots\rangle$ standing for the quantum expectation value, and $k_{\parallel}$ is the wavevector in the 2DES plane.
Equation 1 dictates that SOC effect, e.g.  total spin structure and spin splitting, generated by Rashba and Dresselhaus interactions, is independent of their relative contributions to the total SOC effect. 
The expression in squared brackets acts as an effective SOC coefficient and for a free electron system it produces two Fermi surfaces with opposite spin chiralities.   
By contrast,  in zinc blende GaAs 2DES formed in (001)-plane Rashba and Dresselhaus couplings produce different spin structures. 
The total SOC effect and the resulting band structure depend on the relative contribution of Rashba and Dresselhaus components defined by the details of the confinement potential ~\cite{SOIGaAs, PSH-GaAs2007,PSH-GaAs2009, PSH-GaAs2016}.

\begin{figure}[!thb]
\includegraphics{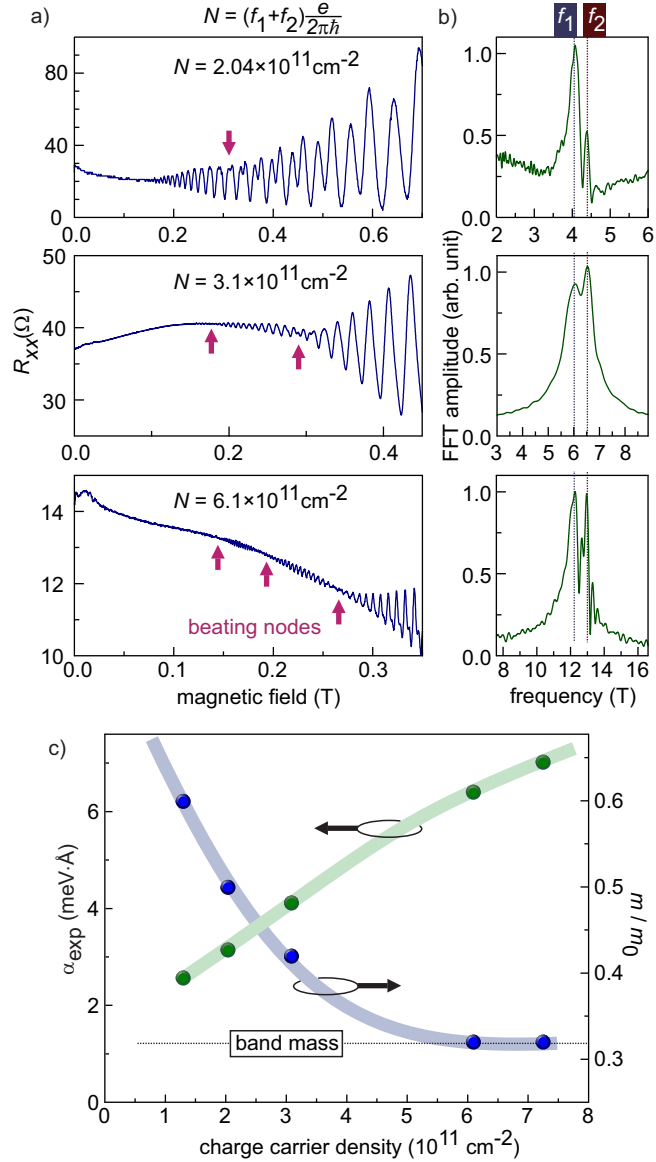}
\caption{\label{Fig2} \textbf{Electron density dependence of spin-orbit coupling effect and Coulomb interaction.} \textbf{a)} In low magnetic field Shubnikov-de Haas effect reveals the beating pattern of quantum oscillations in a wide range of charge carrier density. The arrows indicate the beating nodes. \textbf{b)} Fast Fourier transformation reveals two dominating oscillation frequency components. \textbf{c)} Spin-orbit coupling coefficient (left axis) extracted from the Fourier spectrum shown in panel b) depends on the charge carrier density according to Eq.\,2. The electron effective mass (right axis, $m_0$ is the free electron mass) increases with decreasing electron density and points to Fermi-liquid-like behavior. ZnO is thus a system showing an interplay between SOC effect and Coulomb interaction effect.}
\end{figure}

By performing magnetotransport experiment in ZnO we resolve an SdH beating pattern, which thus points to the presence of at least two Fermi surfaces (see Method Section). 
Examples of beating patterns are shown in Fig.~\ref{Fig2}a.
To identify the size of each Fermi surface we plot in Fig.~\ref{Fig2}b the Fourier transformation spectrum of the SdH signal shown in Fig.~\ref{Fig2}a. 
It clearly visualises the presence of two frequencies labeled $f_\text{1}$ and $f_\text{2}$.
Their relation to the respective Fermi surface areas $A_{1, 2}$ is $f_{1,2}={\hbar}A_{1,2}/{2\pi e}$, where $e$ is the elementary charge and $\hbar$ is the reduced Planck constant. 
We exclude the population of the second subband of the confinement potential to yield two distinct frequencies, 
since it is populated at electron density $N>10^{12}$ $\text{cm}^{-2}$\,\cite{WaveFunctionZnO, TwoSubbandsZnO}. 
Neither the sample inhomogeneity, such as a presence of 2DES areas with distinct charge carrier densities, 
yields two frequencies. It rather results in the smearing of SdH oscillation.

\begin{figure}[!thb]
\includegraphics{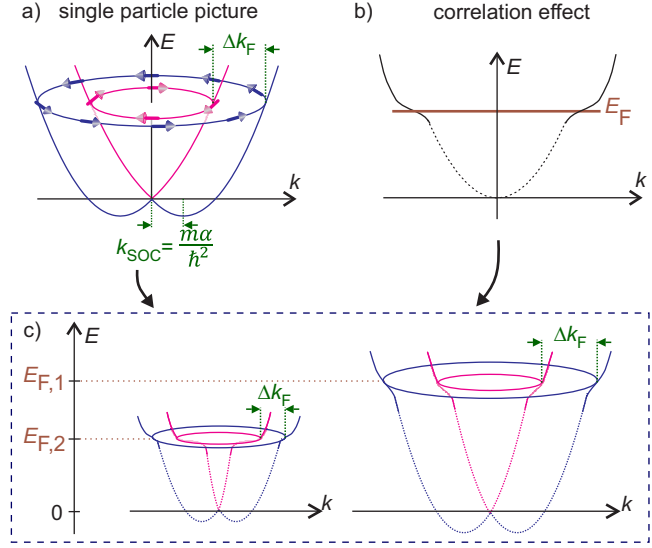}
\caption{\label{Fig3} \textbf{Evolution of spin-orbit split band structure due to Coulomb interaction} 
\textbf{a)} Upon exposing a free electron system to the SOC $H_{\rm SOC}=\alpha (\sigma_{x}k_{y}-\sigma_{y}k_{x})$, the 
energy dispersion becomes $E _{+/-}(k)=\hbar ^2k^2/2m \pm \alpha \, k$.  Here the electron mass $m$ and SOC 
coefficient $\alpha$ are $k$-independent,  and the relation holds $\Delta k_F=2k_\text{\rm SOC}$. 
\textbf{b)} The effect of Coulomb interaction can be thought as flattening of energy dispersion at the Fermi energy. 
\textbf{c)} Schematic representation of  effect of Coulomb interaction on spin-orbit split band structure at two Fermi energy values $E_{F,1}$ and 
$E_{F,2}$. The electron correlation effects are stronger pronounced at smaller Fermi energy. }
\end{figure}

We suggest the formation of two Fermi surfaces  is the result of SOC effect. Figure\,\ref{Fig3}a depicts their realisation according to the Eq.\,1. 
Since the Fermi energy is the same for both surfaces, the difference in their Fermi wavevectors is given by  $\Delta k_F \propto \alpha m$, where $\alpha$ and $m$ are the SOC coefficient and the electron mass at the Fermi surface, respectively.  
This condition allows evaluating the SOC coefficient from the experiment as:
\begin{equation} 
\alpha=\frac{\hbar^2 \Delta n}{m} \sqrt{\frac{\pi}{2N}},
\end{equation} 
where $\Delta n= (f_{1}-f_{2}){e}/{2\pi\hbar}$ and  the total electron density $N=(f_{1}+f_{2}){e}/{2\pi\hbar}$  (Supplementary Note 1). 
Since the  experiment probes the 2DES properties at the Fermi surface,  $m$ and  $\alpha$ can appear renormalised by electron correlation effects.  
In fact, upon lowering the electron density, the effective mass increases (Fig.~\ref{Fig2}c (right axis)) signaling a strong Coulomb interaction. 
It was evaluated  from the temperature dependence of SdH oscillation amplitude (Supplementary Note 2). 
Such  a strong mass enhancement is consistent with our previous studies \cite{Maryenko2012, Kasahara2012, FalsonAPL2015}. 
Now the SOC coefficient can be estimated according to Eq.\,2 with $N-$dependent mass. It is shown in Fig.~\ref{Fig2}c (left axis) and it decreases upon lowering $N$. 
The estimated values are higher than previously reported $\alpha_{\rm exp}=0.7$ meV${\rm \AA}$ measured in the electron spin resonance at high magnetic field~\cite{KozukaSOI}.

Figure~\ref{Fig2}c constitutes the main result of our experimental study -  the decrease in the density leads to a strong enhancement of the electron mass accompanied by a decrease of the SOC coefficient. 
This can be viewed as a variation of electron dispersion shown schematically in Fig.~3. Here Fig.~3a corresponds to the band structure with SOC in the absence of the electron-electron interaction. In accordance with the Fermi liquid theory, the Coulomb interaction enhances the mass shown in Fig.~3b as a flattening of the dispersion at the Fermi surface. 
The combined effect of SOC and Coulomb interaction is visualized in Fig.~3c, where the dispersion curves are shown for two different Fermi energies, so that the split of Fermi surfaces $\Delta k_{F}$ changes substantially with $E_{F},$ 
as suggested by the result of our experiment.

To understand this result in terms of electron-electron interaction we performed perturbation theory calculations using the parameters of 2DES in ZnO (see Supplementary Note 3). 
Although this approach cannot capture the correlation phenomena accurately, it reproduces the tendencies in the behaviour of $m$ and $\alpha$ with the changes in the Coulomb interaction. 
The main result is that this interaction renormalises both SOC and the effective mass parameters, making them dependent on the electron density and particle's momentum. 
As a result, at the Fermi surface, the calculated effective mass grows and the effective SOC decreases with decreasing electron density, as determined by the same density-dependent Fermi energy for both chiralities. 
This is consistent with the observation presented in Fig.~2c, but not in a good quantitative agreement with the experiment since we use a perturbational approach for the strongly correlated 2DES. 
Additionally, we mention that the correlations  can change the relative contribution of atomic orbitals with different angular momentum to the electron Bloch wavefunction, leading to the modification of $\alpha$ and $m$. 
An adequate theory to describe evolution of the electron spectra and SOC  in the strong correlation regime is yet to be developed.

\begin{figure*}[!thb]
\includegraphics{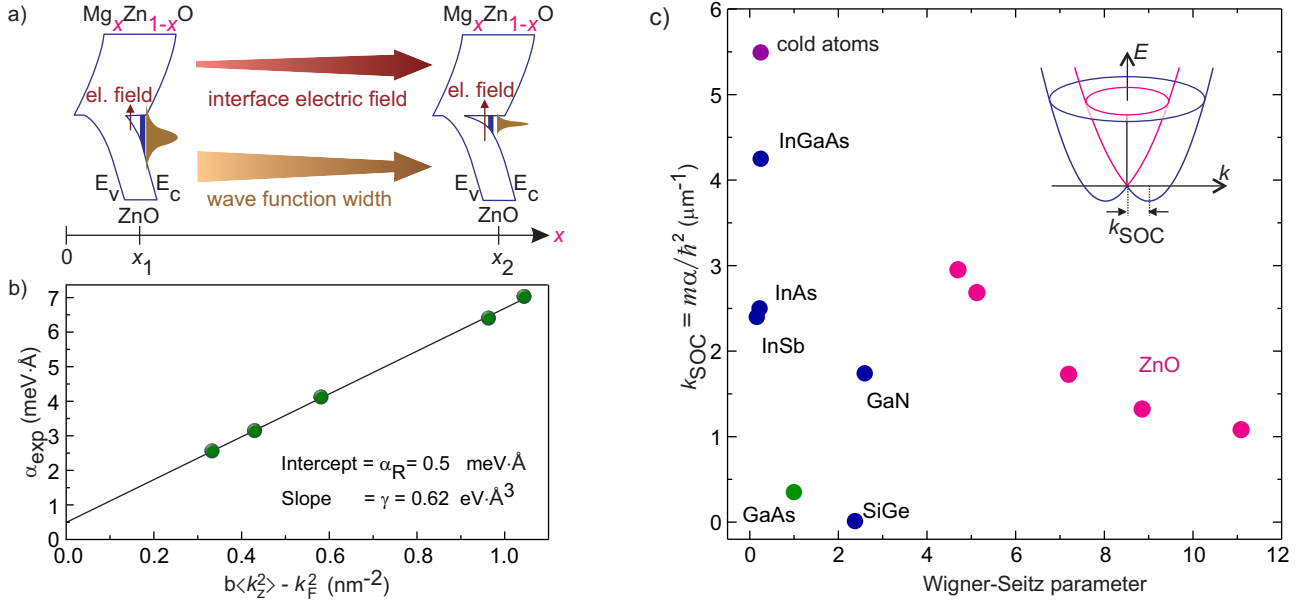}
\caption{\label{Fig4} \textbf{Spin-orbit coupling effect in semiconductor 2D systems}. \textbf{a)} 
Schematic representation of the confinement potential change with  Mg content.  Interfacial electric field increases, while the wavefunction shrinks, as Mg-content increases. 
\textbf{b)} Estimate of  Rashba $\alpha_{R}$ and Dresselhaus $\gamma$ spin-orbit coupling coefficients. Here  $b=3.85$ for our Fermi wavevector range $k_F=\sqrt{2\pi N}$ \cite{SOIWurtzite1}.
\textbf{c)} Comparison of various compounds in terms of spin-orbit coupling strength given by the wavevector  $k_\text{SOC}$ and Coulomb strength given by  the Wigner-Seitz parameter $r_s=1 / \sqrt{\pi N a_B^2} $,  where $a_B=\epsilon\hbar^2/ me^2$ is the Bohr radius in cgs units. For calculating $r_s$  we take an electron density for which typical values of spin-orbit coupling coefficient are reported  (Supplementary Note 7). 
The following references are used for each material. GaAs (electrons)~\cite{SOIGaAs}, GaN/AlGaN~\cite{SOIGaN}, InSb/InAlSb~\cite{SOIInSb}, InGaAs/InAlAs~\cite{SOIInGaAs}, InAs/AlSb~\cite{SOIInAs}, SiGe/Si/SiGe~\cite{SOISiGe} , cold atoms~\cite{ColdAtoms}. }
\end{figure*}

Finally we note that there are other possible origins for mass enhancement at the Fermi energy, which are not directly related to SOC renormalisation.  
Two of them  are related to (1) the role of the vacancies in formation of the band structure or to (2)  the mass renormalisation  by electron-phonon coupling (see Supplementary Notes 4 and 5). 
However, these mechanisms cannot directly explain the observed behaviour of the mass but can contribute to the mass change.

We now turn our attention to the $N-$dependence of $\alpha_{\rm exp}$ presented in Fig.~\ref{Fig2}c. Beside the renormalisation of SOC coefficient (square brackets in Eq.\,2) due to the correlation effects, $\alpha_{\rm exp}$ dependence can reflect the effect of the interface electric field. The scenario is schematically presented in Fig.~\ref{Fig4}a. 
The larger is the Mg content, i.e., the electron density, the larger is the interfacial electric field that can couple to the electron spin, effectively enhancing $\alpha_{\rm exp}$. 
We estimate an interface electric field approximately as 1 mV/${\rm \AA}$ at $N=10^{12}$ cm$^{-2}$. 
Since ZnO is a light large-gap material, this field is not expected to produce an experimentally measurable SOC and thus cannot account for the change of $\alpha_{\rm exp}$.  
However, at a larger Mg content  the wave function width shrinks due to the steeper electron confinement potential (Fig.~\ref{Fig4}a), and thus $\left\langle k_{z}^2\right\rangle$ increases. 
According to Eq.~1 the contribution of the Dresselhaus component to SOC changes linearly with $\left\langle k_z^2\right\rangle$. 
Knowing the wavefunction width from our previous studies ~\cite{WaveFunctionZnO}, we estimate $\left\langle k_z^2\right\rangle$ values for all of our structures  (see Supplementary Note 6)  and plot  in Fig.~\ref{Fig4}b $\alpha_{\rm exp}$ vs $b\left\langle k_{z}^2\right\rangle-k_{F}^{2}$.  
All points fall onto one straight line (in black); its slope defines the Dresselhaus coefficient $\gamma=0.62$ eV\,${\rm \AA}^3$, while the intercept gives the Rashba coefficient $\alpha_R$=0.5 meV\,${\rm \AA}$. 
These coefficients are comparable  with the theoretically estimated  Rashba and Dresselhaus coefficients $\alpha_R^{\rm th}$=1.1 meV\,${\rm \AA}$ and $\gamma^{\rm th}=0.33$ eV\,${\rm \AA}^3$, respectively\,\cite{Cardona1996, SOIWurtzite1}. 
 Although the linear dependence of $\alpha_{\rm exp}$ on $\left\langle k_z^2\right\rangle$  supports our evaluation of $\alpha_{\rm exp}$ according to Eq.\,2 valid for the single particle model, $\alpha_\text{exp}$ contains also the renormalization effect. 
The latter contribution cannot be adequately evaluated due to the deficiency of existing models for analysing the experimental results in the strong electron correlation regime. Rather, $\alpha_{\rm exp}$ can be comprised as a SOC coefficient of quasiparticles.

In Fig.~\ref{Fig4}c we compare  ZnO with other semiconductors hosting high mobility 2DES in terms of Coulomb interaction and SOC strength. 
We take the Wigner-Seitz parameter $r_s$ and the wavevector $k_{\rm SOC}$ to characterize the Coulomb interaction and the SOC strength, respectively. 
Here, ZnO stands out because of its large electron mass, small electron density and a moderate SOC coefficient. 
Since $k_{\rm SOC}\propto \alpha m$, the large mass of ZnO compensates for a moderate  $\alpha$ and makes $k_{\rm SOC}$ comparable to that of InAs, a material known for its large  $\alpha$ and small mass. 
Because of a small mass and a large electron density in InAs, $r_s$ is small, so that the electron correlation effects are not  pronounced there. 
To reduce the electron density in such a system while preserving a  high mobility is challenging.  
Another benchmark system is a 2DES of GaAs, which can host diluted 2DES achieving large $\rs$ values comparable to that of ZnO. 
However, the SOC effect is reported for GaAs system with a large electron density~\cite{SOIGaAs, PSH-GaAs2007, PSH-GaAs2016}. 
The SOC coefficient in GaAs is comparable to that of ZnO,  but its small electron mass yields a small $k_{\rm SOC}$.  
We notice that  two-dimensional holes in GaAs may have a large mass, dilute charge carriers, and a relatively large SOC coefficient.
However, this system is not presented here due to its complicated valence band structure featuring a mixture of heavy and light holes producing system-dependent nonparabolic dispersion, and nonlinear in $k$ spin-orbit coupling. 
As a result the bandstructure and Coulomb interaction effects cannot be unambiguously distinguished ~\cite{GaAsHolesWinkler2002, GaAsHolesHabib2004, GaAsHolesNichele2014}. 
For comparison with other classes of systems, demonstrating SOC effect, but other types of interactions, we also added the typical parameters for cold atoms with $\alpha\sim 5\times 10^{-4}$ meV\,$\rm\AA$~\cite{ColdAtoms}.

The observation of the SOC effect in strong electron correlation regime of ZnO-based 2DES reaches into the unprecedented regime of an interplay between spin-orbit and Coulomb interactions. Experimental results can phenomenologically be understood assuming the existence of quasiparticles with certain charge, mass and chirality similar to electrons 
in single-particle approximation with the parameters strongly depending on electron density.  The theoretical framework for SOC-related effects in the regime of strong correlation is, however, lacking, since the relativistic effect on quasiparticle excitation in such a regime is unknown yet. Our work can be a guideline for establishing basic principles of SOC physics to clarify the transitions from spin-locked chiral excitations in weakly interacting systems to novel quasiparticles in the strongly correlated regime. Consequently, it contributes to the understanding of emergent phenomena in modern spintronics brought about by strong correlations.

\subsection{Method} 
\bf{Experimental details  }\rm The samples are Mg$_x$Zn$_{1-x}$O/ZnO heterostructures grown with molecular beam epitaxy and cut in pieces of about 2\,mm x 2\,mm. The charge carrier density is tuned by changing the Mg-content in Mg$_x$Zn$_{1-x}$O layer.
Indium ohmic contacts are attached at each corner. The structures under study cover an electron density range between $1.7\times10^{11}$ cm$^{-2}$ and $8\times10^{11}$cm$^{-2}$. We used the same growth procedure, substrate and heterostructure handling that were employed in all our previous studies. This gives us a fairly reliable reason to apply the structural characteristics of Mg$_x$Zn$_{1-x}$O/ZnO heterostructures from previous studies to our case.  Each sample is cooled down to base temperature of a dilution refrigerator, which was between 30\,mK and 40\,mK depending on cooling cycle. The magnetotransport is characterized using 4-probe measurement technique and using the lock-in amplifier with an excitation current of 100\,nA. To resolve the beating pattern the magnet sweep rate is set to 5mT/min. The same beating pattern appears at a slower sweep rate, such as for instance 2.5\,mT/min. At higher sweep rates the beating pattern smears out. 

\bf{Band structure calculation }\rm First-principles calculations are performed using a self-consistent Green function method~\cite{Hoffmann2020} within the density functional theory (DFT) in a generalised gradient approximation~\cite{Perdew1996}.  The method is specially designed to study electronic, magnetic, and transport properties of  semi-infinite systems like surfaces and interfaces. Oxygen vacancies, substitutional and anti-site disorder are taken into account within a coherent-potential approximation as it is implemented within the multiple-scattering theory~\cite{Gyorffy1972}.   The band gap size of ZnO and Mg$_x$Zn$_{1-x}$O is strongly underestimated since DFT can not describe correctly excited state properties by construction. However, the behavior of the band gap as a function of layers in a Mg$_x$Zn$_{1-x}$O/ZnO interface should be well mimicked schematically. The crystalline structure of the  Mg$_x$Zn$_{1-x}$O/ZnO interface was adopted from Ref.~\cite{Betancourt2013}.

\textbf{Acknowledgment} This work was supported by the National Science Center in Poland as
a research project No.~DEC-2017/27/B/ST3/02881. AE acknowledges
financial support from the DFG through priority program SPP1666
(Topological Insulators), SFB-TRR227,  and OeAD Grants No. HR 07/2018 and No. PL 03/2018. EYS acknowledges support by the Spanish Ministry of Science and the  
European Regional Development Fund through PGC2018-101355-B-I00  
(MCIU/AEI/ FEDER, UE) and the Basque Country Government through Grant  
No. IT986-16. YK acknowledges support by JST, PRESTO Grant Number JPMJPR1763.

\textbf{Author Contributions:} DM conceived the project.  DM and M. Kawamura performed experiments. M. Kriener and YK contributed to the experiment. DM, M. Kawamura and MSB discussed experiment at the initial stage of the project. AE performed first principles calculations. VKD and EYS provided theoretical support. DM, VKD, EYS and AE wrote the manuscript. M. Kawasaki supervised the project. All authors discussed results and commented on manuscript.


\appendix
\setcounter{equation}{0}
\setcounter{figure}{0}
\renewcommand{\thefigure}{S\arabic{figure}}
\renewcommand{\thetable}{S\arabic{table}}

\section{Supplementary Notes}

\subsection{\bf 1. Evaluation of spin-orbit coupling coefficient $\alpha$ from experimental data}

This section describes the derivation of Eq. (2) in the main text and discuss the possible consequences for the case 
of renormalized mass and SOC coefficient. The derivation relies on the fact that $m$ and $\alpha$ are $k$ wavevector independent 
and are the same for both branches of energy dispersion. 

We consider a free electron system with the spin-orbit coupling, whose Hamiltonian is given by:
\begin{equation}
H_{\rm total}=H_{0}+H_{\rm SOC}=\frac{\hbar^2k^2}{2m}+\alpha (\sigma_{x}k_{y}-\sigma_{y}k_{x}),
\end{equation}
where $m$ is the electron effective mass and $\alpha$ is the spin-orbit coupling coefficient. 
The Hamiltonian has two eigenvalues
\begin{equation}\label{varepsilonpm}
\varepsilon_{\pm}=\frac{\hbar^2k^2}{2m}\pm \alpha k,
\end{equation}
each describing the energy dispersion of the corresponding band, which cross at $k=0$. We call the band described by $\varepsilon_{+}$ ($\varepsilon_{-}$) 
as the inner(outer) band.

Consider the realization where same for both bands Fermi energy $\varepsilon_{F}$ lies above the crossing point at $k=0.$
Then the Fermi wavevectors for the inner and outer bands are:
\begin{eqnarray}
\label{kinner}
k_{F,+} =\frac{m}{\hbar^2}\left[-\alpha+\sqrt{2\pi N \frac{\hbar^4}{m^2} -\alpha^2} \right] \\
\label{kouter}
k_{F,-}=\frac{m}{\hbar^2}\left[\alpha+\sqrt{2\pi N \frac{\hbar^4}{m^2} -\alpha^2} \right],
\end{eqnarray}
where $N$ is the total electron density. 
At $\alpha=0,$ we introduce single $k_{F}\equiv\,k_{F,+}=k_{F,-},$ and the Fermi surface area for each band $A_F\equiv\pi k_{F}^{2}$
is associated with the oscillation frequency $f$ of Shubnikov-de Haas oscillations by the relation:
\begin{equation}
f=\frac{\hbar}{2\pi e}A_F.
\end{equation}

Now, in the experiment at $\alpha\ne\,0$ (see Fig. 2 in the main text) we obtain two frequencies $f_1$ and  $f_2$, which 
are associated with the inner and outer Fermi surfaces respectively.  
The area difference of two Fermi surfaces is:
\begin{equation}
\label{DeltaA1}
\begin{aligned}
\Delta A_{F}= &\pi k_{F,-}^2-\pi k_{F,+}^2=4\alpha\frac{\pi m}{\hbar^2}\sqrt{2\pi N -\frac{m^2}{\hbar^4}\alpha^2}
\end{aligned}
\end{equation}
This relation establishes the connection between the strength of spin-orbit coupling $\alpha$ and the difference of Fermi pocket cross sections. 
On the other side
\begin{equation}
\label{DeltaA2}
\Delta A_F= \frac{2\pi e}{\hbar}\Delta f = 4\pi^2\Delta n. 
\end{equation}
where the notation  $\Delta n\equiv e\Delta f /2\pi \hbar$  is introduced.

Then Eqs. (\ref{DeltaA1}) and (\ref{DeltaA2}) yield:
\begin{eqnarray}
\Delta n= \frac{\alpha m }{\pi \hbar^2}\sqrt{2\pi N -\frac{m^2}{\hbar^4}\alpha^2} .
\end{eqnarray}

Solving this equation one obtains:
\begin{equation}
\alpha^2 = \frac{\hbar^4 \pi}{m^2}\left[ N\pm \sqrt{N^2 -(\Delta n)^2}  \right].
\end{equation}
Since both solutions satisfy the request $\alpha^2 >0$ and only one $\alpha$ can be the solution, 
we consider another restriction. If $\alpha=0$, there should be only one oscillation frequency and 
thus $\Delta n=0$. Then  only one solution remains. 
\begin{equation}
\begin{aligned}
\alpha^{2}=&\frac{\hbar^4 \pi}{m^2}\left[ N - \sqrt{N^2 -(\Delta n)^2}  \right]  
 = \frac{\hbar^4 \pi}{m^2}\frac{N^2 - (N^2 -(\Delta n)^2) } {N + \sqrt{N^2 -(\Delta n)^2} } .
\end{aligned}
\end{equation}
 It follows:
\begin{equation}
 \alpha=\frac{\hbar^2}{m}\Delta n \frac{\sqrt{\pi}}{\sqrt{N+\sqrt{N^2-(\Delta n)^2}}}.
\end{equation}
At $N\gg \left(m\alpha^{2}/\hbar^{2}\right)^2$ we have $\varepsilon_{F}\approx\pi\hbar^{2}N/m$ and $\Delta n \ll N.$  Thus, we obtain Eq. (2) of the main text:
\begin{equation}
\label{Eq2}
\alpha=\frac{\hbar^2}{m}\Delta n \sqrt{ \frac{\pi}{2N}  }.
\end{equation}

\subsection{\bf 2. Estimation of electron effective mass}
We demonstrate the estimation of the electron effective mass $m$ from the Shubnikov-de Haas oscillations 
on the example of one Mg$_x$Zn$_{1-x}$O/ZnO structure, which is shown in the middle panel of Fig.2a in the main text. 
We consider that the oscillation of magnetoresistance follows Lifshitz-Kosevich formalism \cite{LKPaper}. 
We also consider only the leading term in Taylor series of Lifshitz-Kosevich approach:
\begin{eqnarray} 
\label{LK}
&& \frac{\Delta R_{xx}}{R_{xx}}=\frac{4XT}{\sinh{(XT)}}  \exp(-\pi/\omega_c\tau) \\
&& X=\frac{2\pi^2k_B}{\hbar \omega_c}, \nonumber
\end{eqnarray}
where $T$ is the temperature, $\omega_{c}=eB/m$ is the cyclotron frequency at field $B$, and $k_B$ is the Boltzmann constant. 
\begin{figure}[!thb]
\includegraphics{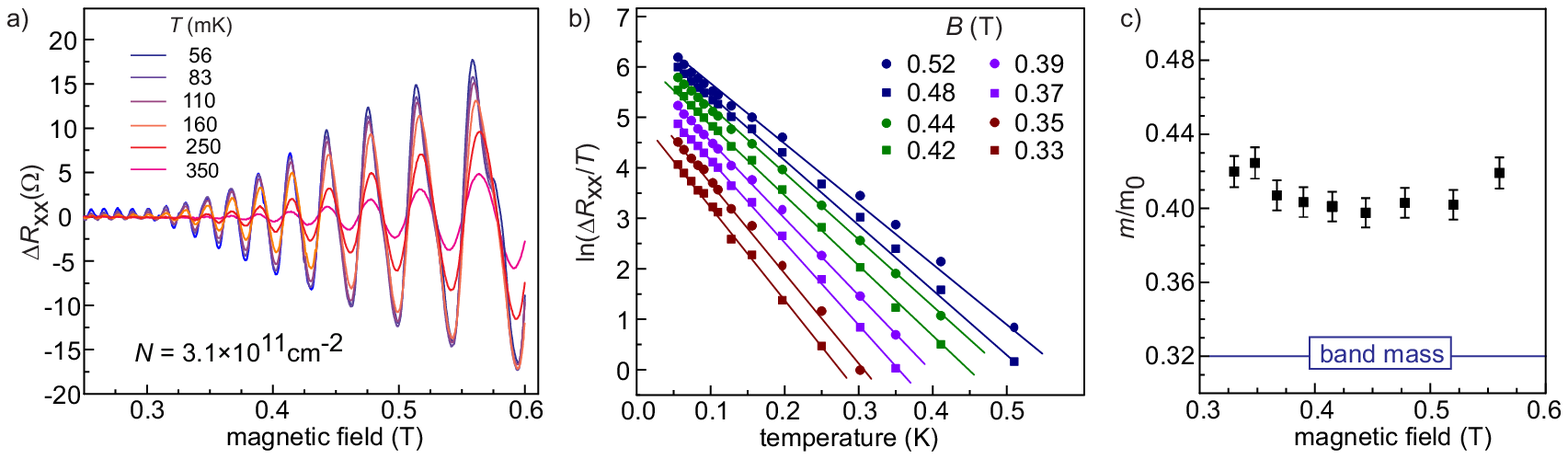}
\caption{\label{MassSI} \textbf{Estimate of electron mass} 
\textbf{a)} Oscillatory part of the magnetoresistance at various temperatures
\textbf{b)} Mass analysis according to Lifshitz-Kosevich approximation. The slope at a given magnetic field defines the electron effective mass.
\textbf{c)} Mass as a function of magnetic field. The mass enhancement is attributed to the electron correlation effects.}
\end{figure}

Figure~\ref{MassSI}a shows the development of oscillating part of the magnetoresistance with temperature, while panel 
(b) demonstrates the mass analysis according to Eq.~(\ref{LK}). The mass is estimated for several field values 
and is plotted in Fig.~\ref{MassSI}c. At $N=3.1\times 10^{11}$ cm$^{-2}$, it amounts to $m$=0.41$m_{0}$, where $m_{0}$ is the free electron mass,
enhanced compared to the bulk value of 0.32$m_0$. This enhancement is attributed to the correlation effects.


\subsection{\bf 3. Renormalization of the effective mass and spin-orbit coupling by electron-electron interactions}  

Here we present a model for renormalization of the electron mass and spin-orbit coupling and their changes with the charge carrier density
using the analysis based on the perturbation theory. The Hamiltonian of two-dimensional (2D) electron system with the Rashba SO coupling and 
electron-electron interaction is written as the sum of two terms
\begin{equation}
\label{H7}
H=H_0+H_{\rm int},
\end{equation}
where
\begin{equation}\label{H0Hint}
H_{0}=\psi ^\dag ({\bf r})\Big[ -\frac{\hbar ^2\Delta }{2m_{b}}
-i\alpha_{b}\, \hat{\bf z}\cdot (\bsig \times {\bm \nabla})\Big] \, \psi ({\bf r}), \qquad 
H_{\rm int}=v({\bf r}-{\bf r'})\, \big[ \psi ^\dag({\bf r})\, \psi ({\bf r})\big]
\big[ \psi ^\dag ({\bf r'})\, \psi ({\bf r'})\big]. 
\end{equation}
Here $m_{b}$ and $\alpha_{b}$ is the bare (interaction-independent) electron effective mass and the Rashba coupling constant, 
respectively, $\hat{\bf z}$ is the unit vector perpendicular to the $x$-$y$ plane, $\psi ^\dag ({\bf r})$ 
and $\psi ({\bf r})$ are the spinor field operators, and the function $v({\bf r})$ with ${\bf r}=(x,y)$ 
describes Coulomb interaction of electrons at a distance $r$.
We assume that Coulomb interaction is screened, and its Fourier component is \cite{Ando}
\begin{eqnarray}
\label{vq}
v(q)=\frac{2\pi e^2}{\epsilon\, (q+\kappa )}, 
\end{eqnarray}
where $\epsilon$ is the dielectric constant and $\kappa $ is the inverse screening length. 
%
The Schr\"odinger equation with Hamiltonian $H_{0}$ gives the two-band dispersion 
$\varepsilon_{\lambda}(k)=\varepsilon(k)\pm \alpha_{b}k$, where $\varepsilon(k)=\hbar^{2}k^{2}/2m_{b}$
and $\lambda=+(-)$ is the chirality index corresponding to the spin-related branch of the spectrum (cf. Eq. (\ref{varepsilonpm})).

\begin{figure}
	\centering
		\includegraphics{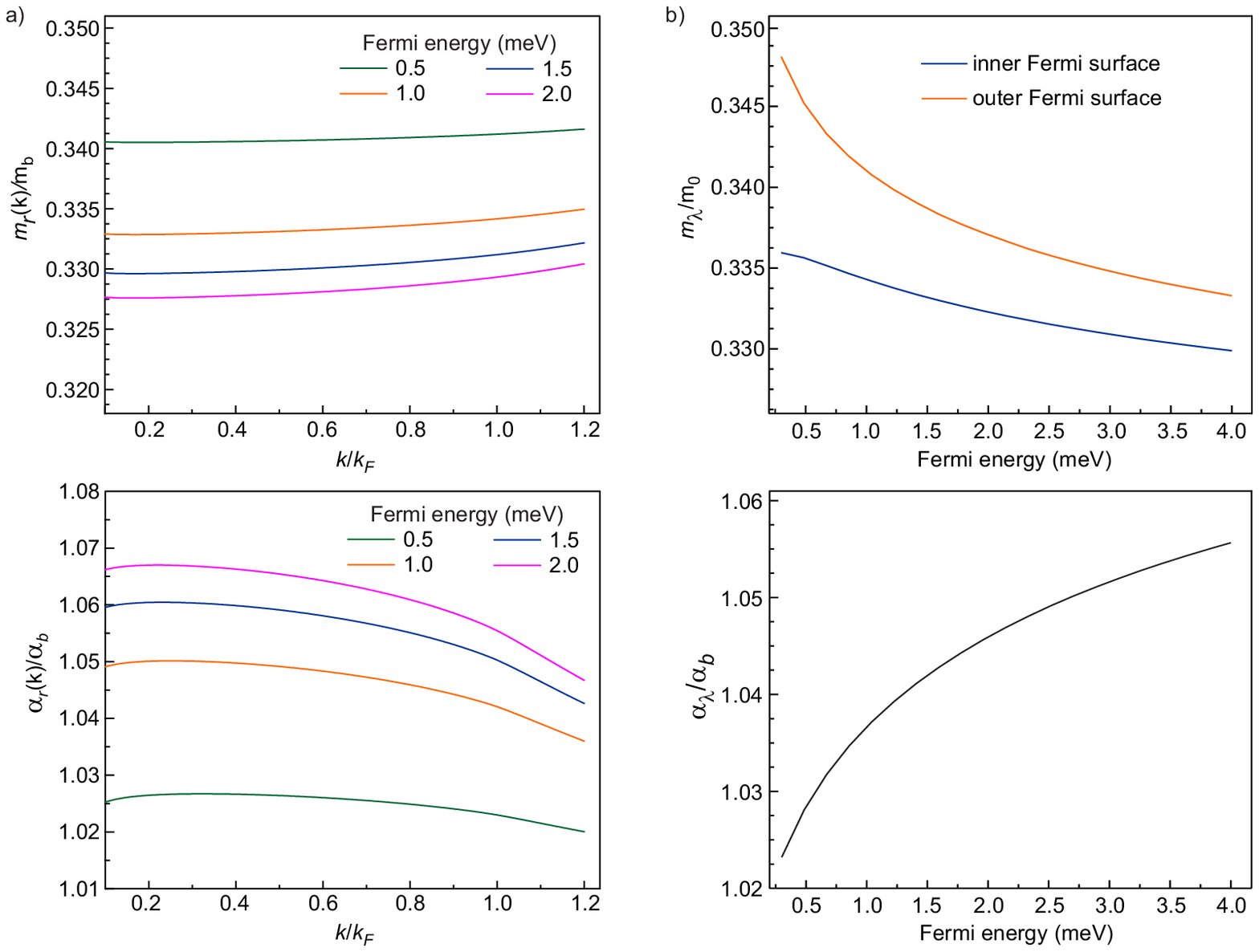}
	\caption{a) Renormalized by electron-electron interaction 
	parameters $m_{r}(k)=m_{b}\chi_{1}^{-1}(k)$ and $\alpha_{r}(k)=\alpha_{b}\chi_{2}(k)$ in the self-consistent approach of Hamiltonian 
	(\ref{H0Hint}) as a function of $k/k_{F}$ for different chemical potentials $\mu$. 
	Here $m_{0}$ is the mass of free electron. b) Effective mass $m_{\lambda}$ (calculated with Eq. (\ref{mstar})) and Rashba coupling $\alpha_{\lambda}$ 
	at the Fermi surfaces, 	$k=k_{F,+},k_{F,-},$ as a function of $\mu$. The chirality index $\lambda=(+)$ describes the inner Fermi surface, while $\lambda=(-)$ 
	describes the outer Fermi surface. Renormalised SOC coefficient $\alpha_{r}$ dependence on $\mu$ is the same for both Fermi surfaces.}
	\label{FigSI}
\end{figure}

The exchange and Hartree  diagrams \cite{Abrikosov} for the Coulomb interaction yield the following contributions to the self energy of electrons:
\begin{eqnarray}
\label{Sigma1}
&&\hat{\Sigma }_{\rm xc}({\bf k})
=i\int \frac{d\varepsilon }{2\pi }\frac{d^2{k'}}{(2\pi )^2}\,
v({\bf k-k'})\, \hat{G}_0(\varepsilon ,{\bf k'}),
\\
\label{Sigma2}
&&\hat{\Sigma }_{\rm H}({\bf k})
=-iv(0)\, {\rm tr} \int \frac{d\varepsilon }{2\pi }
\frac{d^2{k'}}{(2\pi )^2}\, \hat{G}_0(\varepsilon ,{\bf k'}),
\end{eqnarray}
where $\hat{G}_0(\varepsilon ,{\bf k})$ is the $2\times 2$ matrix Green's function of free electron corresponding to Hamiltonian $H_{0}$.
After integrating the Green's function over $\varepsilon $ we obtain
\begin{equation}\label{Sigma2int}
\int \frac{d\varepsilon }{2\pi }\, \hat{G}_0({\bf k'},\varepsilon )
=\frac{i}{2} \Big\{ \theta [\mu -\varepsilon_{+}(k')]+\theta [\mu -\varepsilon_{-}(k')]\Big\}\left(1+\hat{\bf z}\cdot (\bsig \times{\bf n}_{\bf k'})\right)
\end{equation}
where $\mu $ is the chemical potential (at zero-temperature equivalent to the Fermi energy), 
$\hat{\bf n}_{\bf k}$ is the unit vector along ${\bf k}$ and $\theta (x)$ is the Heaviside step function.
Substituting (\ref{vq}) and (\ref{Sigma2int}) into Eqs.~(\ref{Sigma1}) and (\ref{Sigma2}) we find
\begin{eqnarray}
\label{Sigmaxc}
\hat{\Sigma }_{\rm xc}({\bf k})
=-\frac{e^2}{2\pi \epsilon} \int _0^{\pi }d\varphi \int _0^{k_{F,+}}
\frac{k'dk'}{\zeta ({\bf k,k'})+\kappa }-\frac{e^2\hat{\bf z}\cdot (\bsig \times {\bf n}_{\bf k})}
{2\pi \epsilon}
\int _0^{\pi } \cos \varphi d\varphi 
\int _0^{k_{F+}}\frac{k'dk'}{\zeta ({\bf k,k'})+\kappa }
\nonumber \\
-\frac{e^2}{2\pi \epsilon} \int _0^{\pi }d\varphi \int _0^{k_{F-}}
\frac{k'dk'}{\zeta ({\bf k,k'})+\kappa }+\frac{e^2\hat{\bf z}\cdot (\bsig \times {\bf n}_{\bf k})}{2\pi \epsilon}
\int _0^{\pi }\cos \varphi  d\varphi \int _0^{k_{F-}}\frac{k'dk'}{\zeta ({\bf k,k'})+\kappa }\, ,
\end{eqnarray}
\begin{eqnarray}
\label{SigmaH}
\hat{\Sigma }_{\rm H}({\bf k})=\frac{e^2(k_{F+}^2+k_{F-}^2)}{2\epsilon\kappa },
\end{eqnarray}
where the Fermi wavevectors $k_{F,\lambda}$ are determined by common chemical potential $\varepsilon_{\lambda}(k_{F,\lambda})=\mu $ 
(cf. Eqs. (\ref{kinner}), (\ref{kouter}),
$\zeta ({\bf k,k'})=\sqrt{k^2+{k'}^2-2kk'\cos \varphi } $, and
$\varphi $ is the angle between vectors ${\bf k}$ and ${\bf k'}$.
Note that (\ref{Sigmaxc}) contains the terms in form of spin-orbit interaction. 
The real part of self energy determines correction to the electron spectrum due to the interactions.

The Hartree contribution does not depend on $k$ and, correspondingly, does not affect the energy structure. 
In our calculations we omit $k$-independent terms in (\ref{SigmaH}) assuming that they only lead to a uniform
shift of all electron energies.  
The $k$-dependent contribution from the exchange diagrams can be written as
\begin{eqnarray}
\label{Sigmaxck}
\hat{\Sigma }_{\rm xc}({\bf k})=\hat{\Sigma }_{\rm xc}^{(1)}(k)
+\hat{\bf z}\cdot (\bsig \times {\bf n}_{\bf k})\, 
\hat{\Sigma }_{\rm xc}^{(2)}(k),
\end{eqnarray}
where the first and second terms stand for corresponding corrections to the Hamiltonian without 
interaction, $\hat{H}_{0{\bf k}}=\varepsilon(k)+\alpha_{b}\hat{\bf z}\cdot (\bsig \times {\bf k})$. 
Note that in the limit $\alpha_{b}\rightarrow\,0,$ the first term, $\hat{\Sigma }_{\rm xc}^{(1)}(k)$ tends 
to a constant (\ref{Sigmaxc}) while $\hat{\Sigma }_{\rm xc}^{(2)}(k)$ vanishes since it is nonzero solely
due to different Fermi momenta $\hbar\,k_{F,\lambda}$ in Eq. (\ref{Sigmaxc}). 

These results are obtained in the first-order perturbation theory. Within 
this approach any higher order corrections to the spectrum should be small. 
To get more realistic results we use the self-consistent approach taking the Hamiltonian 
\begin{eqnarray}
\label{Hk}
H_{\bf k}=\varepsilon(k)\chi_{1}(k)+\alpha_{b}\,\hat{\bf z}\cdot (\bsig \times {\bf k})\, \chi_{2}(k),
\end{eqnarray}
where $\chi_{1,2}(k)$ are some unknown functions to be determined self-consistently. 
They can be presented as $\chi_{1}(k)=m_{b}/m_{r}(k)$ and $\chi_{2}(k)=\alpha_{r}(k)/\alpha$, where $m_{r}(k)$ and $\alpha_{r}(k)$ 
are the renormalized $k$-dependent parameters. 
Then in the first-order perturbation theory we get 
\begin{equation}
\label{f1kf2k}
\chi_{1}(k)\simeq 1+\frac{\Sigma _{\rm xc}^{(1)}(k)}{\varepsilon(k)}; \qquad
\chi_{2}(k)\simeq 1+\frac{\Sigma_{\rm xc}^{(2)}(k)}{\alpha_{b}k}, 
\end{equation}
and in agreement  with the perturbation approach, $\chi_{1}(k)$  and $\chi_{2}(k)$ are close to 1. 

In frame of self-consistent approach we calculate the $\Sigma _{\rm xc}({\bf k})-$dependent Green's function 
with the self energy calculated with full Green's function $\hat{G}({\bf k},\varepsilon )$, which takes 
into account interaction-induced nonparabolicity of the spectrum. Thus, in the previous equations we
substitute $\varepsilon(k)\to \varepsilon(k)\chi_{1}(k)$ and $\alpha_{b}\to\alpha_{b}\chi_{2}(k)$. Correspondingly, we get 
the renormalized spectrum $\tilde{\varepsilon}_{\lambda}(k)=\varepsilon(k)\chi_{1}(k)\pm\alpha_{b}k\,\chi_{2}(k),$
where the Fermi wavevectors in each spin-related subband are determined as the solution of equation of common chemical 
potential $\mu$ as $\tilde{\varepsilon}_{\lambda}(k_{F,\lambda})=\mu$.

We calculated the self energy and the spectrum by iterations starting with the 2D Fermi-gas realization $\chi_{1}(k)=\chi_{2}(k)=1,$ 
using the parameters: $m_{b}=0.35\,m_{0}$, $\epsilon=8.5$, and $\alpha_{b}=2$ meV${\rm \AA}.$ For the 
parameter $\kappa$ we use an approximation, which takes into account the effect of 
Coulomb interaction on the screening. At $r_{s}\ll 1$ the value of $\kappa $ can be determined by 
the random phase approximation (RPA) as $\kappa_{0}=-(2\pi e^2/\epsilon)\, \Pi (\mathbf{0},0)=2\pi e^2\rho(\mu)/\epsilon$, 
where $\Pi ({\bf q},\omega )$ is the polarization operator of two-dimensional electron gas 
and $\rho (\mu )$ is the density of states at Fermi level. Since the SO coupling 
is relatively weak, for $\mu \gg m_{b}\alpha_{b}^2/\hbar^2$ the chosen numerical parameters yield
$\kappa_{0}\simeq 1.27\times 10^{7}$ cm$^{-1}$. 

We assume that for $r_s\gg 1$ the value of $\kappa $ is essentially renormalized by electron-electron interaction. 
For example, within the Landau theory for three-dimensional electron system one can get $\kappa \sim 1/r_{s}$ for $r_{s}\gg 1$. 
In the 2D case one can expect, based on the calculations of the 
interaction functions in the Landau theory for strongly screened potential (\ref{vq}), that at $r_{s}\gg 1,$ 
the parameter $\kappa$ tends to a $r_{s}-$independent constant. 
Thus, in our calculations we use an interpolation formula
\begin{equation}
\label{kappa}
\kappa =\frac{\kappa_0}{1+\gamma r_{s}(1+r_{s})^{-1}},
\end{equation}         
where $\gamma$ is a constant and take  $\gamma=1$ as an example.  

Figure \ref{FigSI}a shows the behavior of functions $m_{r}(k)=m_{b}\chi_{1}^{-1}(k)$ and $\alpha_{r}(k)=\alpha_{b}\chi_{2}(k)$, 
which are $k$-dependent renormalized parameters of the mass and Rashba coupling. The function $k\alpha_{r}(k)$ determines the band splitting at each $k$. 
The main result presented in Fig. \ref{FigSI}a is that the electron-electron interaction-induced renormalization of 
$\alpha_{r}(k)$ and $m_{r}(k)$ makes them $k-$dependent and instead of bare constants $\alpha_{b}$ and $m_{b}$ we obtain functions $\alpha_{r}(k)$ and $m_{r}(k)$. 
Correspondingly, the interactions modify the shape of dispersion curves to $\varepsilon_{\lambda}(k)=\hbar^{2}k^{2}/2m_{r}(k)\pm \alpha_{r}(k)\,k$. 
Moreover,  Fig.~\ref{FigSI}a shows that all these functions depend on the density of electrons, which makes mass $m_{r}(k)$ 
smaller and $\alpha_{r}(k)$ larger with increasing electron density. As we see, $m_{r}(k)$ 
is lower than $m_{b}$ and $\alpha_{r}(k)$ is larger than $\alpha_{b}$ for any $k$ with this effect being stronger for larger $\mu $.
With increasing $k$ the function $m_{r}(k)$ grows to the bare value $m_{b}$, whereas $\alpha_{r}(k)$ 
decreases to the bare $\alpha_{b}$.  

In addition, we find the renormalized electron cyclotron effective mass $m_{\lambda}(k)$ by using the standard definition 
\begin{eqnarray}
\label{mstar}
\frac1{m_{\lambda}(k)}=\frac{1}{\hbar^{2}k}\, \frac{d\varepsilon_{\lambda}(k)}{dk},
\end{eqnarray}
defined for each branch of the spectrum, $\varepsilon_{\lambda}(k)$. 

Taking the effective mass $m_{\lambda}(k)$ and parameter $\alpha_{r}(k)$ at the Fermi surfaces ($k=k_{F,\lambda}$), corresponding to given
value of $\mu$, we obtain the dependence of redefined quantities $m_{\lambda}\equiv\,m_{\lambda}(k_{F,\lambda})$ 
and $\alpha_{\lambda}\equiv\alpha_{r}(k_{F,\lambda})$ on the chemical potential. 
These dependences  are presented in Fig.~\ref{FigSI}b, where $m_{\lambda}$ decreases 
whereas $\alpha_{\lambda}$ increases with the increase in the chemical potential and, therefore, in the electron density.


Earlier, the effect of electron-electron interaction in spin-orbit coupling has been 
considered in Ref.~\cite{Raikh1999}. It was found that in frame of the RPA the interaction 
reduces effective mass and enhances spin-orbit coupling. This is in agreement with our calculations 
(see Fig. \ref{FigSI}b, where $m_{\lambda}<m_{b}$ and $\alpha_{\lambda}>\alpha_{b}$ for all $\mu$). However, as we see, the variation of 
the mass $m_{\lambda}$ and $\alpha_{\lambda}$  
with chemical potential $\mu$ corresponds to increase of $m_{\lambda}$ and decrease of $\alpha_{\lambda}$ at smaller density of electrons.

Qualitatively, theoretical results are in some agreement with the experiment (see Fig.~\ref{FigSI}b) but quantitatively 
the disagreement is rather strong. The main reason is that in strong-interaction regime the existing theory cannot describe 
the electron spectrum but allows to relate certain interaction parameters with observable quantities.

\subsection{\bf 4. Defects at Mg$_{x}$Zn$_{1-x}$O/ZnO interfaces }

\begin{figure}[!thb]
\includegraphics{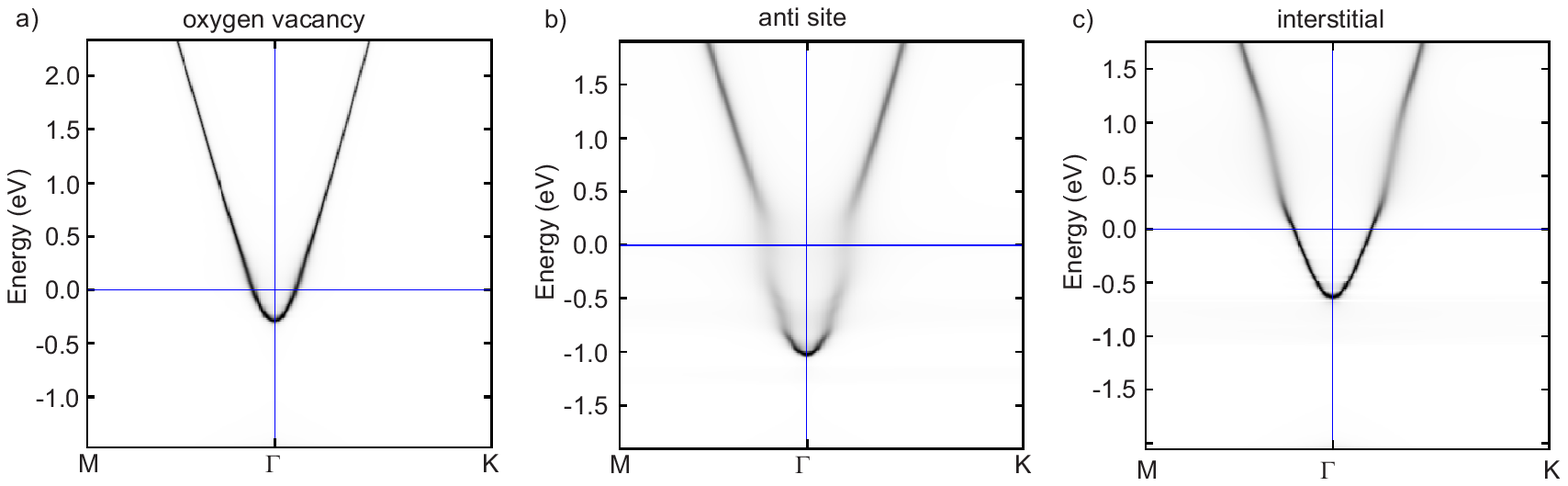}
\caption{\label{defects} Spectral function of the
  Mg$_{0.05}$Zn$_{0.95}$O in presence of defects (0.1\%): (a) oxygen
vacancies; (b) Zn-O anti-site defects; (c)  oxygen interstitial defects.}
\end{figure}

Various defects such as oxygen vacancies, anti-site or interstitial
defects can lead to a strong change of the effective band mass at the
Fermi energy. To demonstrate this fact, we simulated these types of
defects at the Mg$_{0.05}$Zn$_{0.95}$O using a coherent
potential approximation as it is implemented within the multiple
scattering theory for semi-infinite systems~\cite{Hoffmann2020}. The
concentration of impurities was fixed in our simulations to be
0.1\%. The results are presented in Fig.\ref{defects}, which show the
spectral function corresponding to the conductance band in
Mg$_{0.05}$Zn$_{0.95}$O for three types of defects: (i) oxygen
vacancies (Fig.~\ref{defects}(a)); (ii) Zn-O anti-site defects
(Fig.~\ref{defects}(b)); (iii) oxygen interstitial defects
(Fig.~\ref{defects}(c)). All types of considered defects demonstrate a
significant enhancement of the effective band mass in the vicinity of
the Fermi energy. Our simulations should only mimic the impact of
defects on the conductance band, however we have no information about
the real defect structure at Mg$_{x}$Zn$_{1-x}$O/ZnO interfaces. In accordance with
our previous studies, all these defect type are energetically possible
in the ground state. 

\subsection{\bf 5. Mass renormalization due to piezoelectric electron-phonon coupling at the Mg$_{x}$Zn$_{1-x}$O/ZnO interface}

Since ZnO is a strong piezoelectric where electron-phonon coupling
modifies the electron effective mass measured in the cyclotron resonance
experiments in bulk crystal \cite{Mahan1964}, it would be of interest to see the effect of this coupling
in two-dimensional electron systems at the Mg$_{x}$Zn$_{1-x}$O/ZnO
interface. Here electrons interact with two kinds of acoustic phonons, the longitudinal $(l)$ and the  
shear $(s)$ ones. Taking into account that the speed of the lower-frequency shear sound in ZnO
is $s_{s}=2.1\times 10^{5}$ cm/s, the typical phonon energy at the wave
vector $q=10^{6}$ cm$^{-1},$ corresponding to the Fermi momentum at electron concentration $\sim 10^{11}$ cm$^{-2}$
is $\hbar s_{s}q\sim 0.1$ meV is too high for the 
phonon to be excited at experimental temperature 40 mK. For this
reason, a quantum single-phonon perturbation theory is sufficient for calculation of the 
renormalized effective mass. Here we will use the approach proposed by
Hutson \cite{Hutson1961} and Mahan and Hopfield \cite{Mahan1964} to evaluate the corresponding renormalization of 
the electron mass.

The electron-phonon coupling Hamiltonian in terms of electron ($a_{\mathbf{k+q%
}}^{\dag },a_{\mathbf{k}}$) and phonon ($b_{-\mathbf{q}}^{\dag },b_{\mathbf{q}}$) creation and annihilation operators  has the form: 
\begin{equation}
H_{\rm e-ph}=\sqrt{\frac{1}{V}}\sum_{\nu,\mathbf{q}}\frac{\widetilde{e}\sqrt{\hbar}}{\sqrt{\rho s_{\nu
}q}}f_{\nu }\left( \mathbf{q}\right) \left( b_{-\mathbf{q}}^{\dag }+b_{%
\mathbf{q}}\right) a_{\mathbf{k+q}}^{\dag }a_{\mathbf{k}}, 
\end{equation}
where index $\mathbf{\nu }$ denotes the phonon mode ($\mathbf{\nu =}l,$ longitudinal 
and $\mathbf{\nu =}s,$ shear) with velocity $s_{\nu}$ and
wavevector $\mathbf{q,}$ $f_{\mathbf{\nu }}\left( \mathbf{q}\right) $ includes
the strength of piezocoupling, $V$ is the crystal volume, $\rho $ is the
crystal density, and the effective charge $\widetilde{e}=e/\epsilon.$ Here
and below we neglect the change in the phonon properties at the Mg$_{x}$Zn$_{1-x}$/ZnO interface. Neglecting 
small phonon frequency \cite{Mahan1964}, we use perturbation theory for the $k$-dependent
energy shift $\delta\varepsilon_{\rm e-ph}(k)$:
\begin{equation}
\delta \varepsilon_{\rm e-ph}(k)=
\sum_{\nu}\left\langle f_{\nu}^{2}\left(\mathbf{q}\right)\right\rangle\frac{\widetilde{e}^{2}}{\rho s_{%
\mathbf{\nu }}}\frac{\hbar}{\left(2\pi\right)^{3}}\int_{0}^{Q}qdq\int_{0}^{\pi}\sin \theta d\theta \int_{0}^{2\pi }\frac{d\phi }{\varepsilon(\mathbf{k})-\varepsilon(\mathbf{k+q}_{\Vert })},
\end{equation}%
where $Q$ is the cutoff wave vector due to the finite width of the electron wavefunction 
at the interface, and $\left\langle f_{\nu}^{2}\left( \mathbf{q}\right)\right\rangle$ stands for the averaging over directions of 
$\mathbf{q.}$ The in-plane phonon wavevector is given by: $q_{\Vert }=q\sin \theta >0$ (where $\theta$ is the corresponding polar angle), therefore 
\begin{equation}
\varepsilon(\mathbf{k})-\varepsilon(\mathbf{k+q}_{\Vert})=-\frac{\hbar^{2}}{2m_{b}}\left(q_{\Vert}^{2}+2kq_{\Vert }\cos \phi \right), 
\end{equation}
where $\phi$ is the azimuthal angle. 

First integrating over $\phi$ we obtain  
\begin{equation}
\int_{0}^{2\pi}\frac{d\phi}{\varepsilon(\mathbf{k})-\varepsilon(\mathbf{k+q}_{\Vert})}=-\frac{2m_{b}}{q_{\Vert}\hbar^{2}}
\frac{2\pi}{\sqrt{q_{\Vert}^{2}-\left(2k\right) ^{2}}}
\end{equation}
and note that we need $q_{\Vert }>2k$ to get a nonzero integral and,
therefore, $q\geq 2k$.  
Further integration over the polar angle at $q\gg 2k$ yields: 
\begin{equation}
\int_{0}^{\pi}\frac{d\theta}{\sqrt{\sin^{2}\theta -\left( 2k/q\right) ^{2}%
}}=\int_{\arcsin (2k/q)}^{\pi -\arcsin (2k/q)}\frac{d\theta }{\sqrt{\sin
^{2}\theta -\left( 2k/q\right) ^{2}}}=\left( 4\ln 2\right) \ln \frac{q}{2k}. 
\end{equation}
In the large $q\gg 2k$ limit we obtain further%
\begin{equation}
\int_{2k}^{Q}\frac{dq}{q}\ln \frac{q}{2k}=\frac{1}{2}\ln^{2}\left(\frac{2k}{Q}\right). 
\end{equation}



It is convenient to introduce the Fermi velocity $v_{F}$ and to write, as in Eq. (\ref{mstar}),
the renormalized mass at the Fermi surface, $m$, as:  
\begin{equation}
\frac{m_{b}}{m}=1+\frac{2\ln\,2}{\pi^{2}}\sum_{\nu}\frac{\left\langle
f_{\nu }^{2}\left( \mathbf{q}\right) \right\rangle}{\epsilon \rho s_{\nu
}^{2}}\frac{e^{2}}{\epsilon \hbar v_{F}}\frac{s_{\nu }}{v_{F}}\ln \left( 
\frac{Q}{2k_{F}}\right),
\end{equation}
where $k_{F}$ is the Fermi wavevector, and $\ln\left(2k_{F}/Q\right)<0$ by requirement of the narrow density
distribution along the $z-$axis. 
Note that here $e^{2}/\epsilon\hbar$ is the excitonic electron velocity ($\sim 3\times 10^{7}$ cm/s)
with $e^{2}/\epsilon \hbar v_{F}\gg 1.$ 

Using the symmetry analysis of Hutson \cite{Hutson1961}, we obtain that at $Q/2k_{F}\gg 1,$ 
 the main contribution to the mass renormalization is due to the longitudinal
phonons $(\nu=l)$  and the angular averaging $\langle f_{l}^{2}\left( \mathbf{q}%
\right)\rangle $ can be easily performed in this case. For
three-dimensional electrons the values $\langle f_{l}^{2}\left( 
\mathbf{q}\right)\rangle/\epsilon\rho s_{\nu }^{2}$ are given by
the electromechanical coefficient $\left( K_{l}^{2}\right) _{\text{av}}=0.012.$ Taking into account that
in 2DES electron mainly interact with the phonons propagating along the $%
z-$axis, we obtain $\left\langle f_{l}^{2}\left( \mathbf{q}\right)
\right\rangle /\epsilon \rho s_{l}^{2}\approx 10\left( K_{l}^{2}\right)_{\rm av}.$ 
For the given material parameters $s_{l}=5\times 10^{5}$ cm/s 
and typical Fermi velocity $v_{F}=3\times 10^{6}$ cm/s  we obtain as a
result the difference $|m_{b}/m -1|$ less then $0.1$
and, more important, $m$ increases with increasing the Fermi
momentum. This is in agreement with the result of Mahan and Hopfield \cite{Mahan1964} and in
contrast to the experimental observation presented in the main text.



\subsection{\bf 6. Electron wave function width at the Mg$_{x}$Zn$_{1-x}$O/ZnO interface}

The section describes the calculation of $\langle k_{z}^{2}\rangle$ for analysing the spin-orbit coupling effect 
shown in Fig.\,4b of the main text.
\begin{figure}[!thb]
\includegraphics{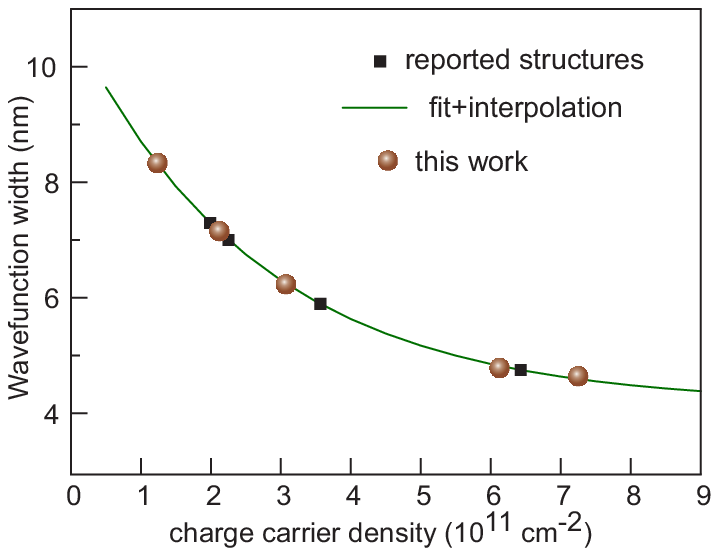}
\caption{\label{SI1} \textbf{Width of wavefunction}. }
\end{figure}
Electron wavefunction at the interface with a triangular-like confinement potential can be presented in the Fang-Howard form as \cite{Ando}:
\begin{equation}
\psi(z)=\frac{\xi^{3/2}}{\sqrt{2}}z e^{-\xi z/2},
\end{equation}
where $\xi$ defines the extension perpendicular to the interface, with the corresponding
expectation values:
\begin{equation}
\langle k_{z}^{2}\rangle \equiv -\int_{0}^{\infty}\psi(z)\psi^{\prime\prime}(z)dz = \frac{\xi^{2}}{4}, \qquad
\langle {z}^{2}\rangle \equiv \int_{0}^{\infty}z^{2}\psi^{2}(z)dz = \frac{12}{\xi^{2}}. 
\end{equation}
The energy minimization yields the relation between $\langle k_{z}^{2}\rangle$ and the electric field $E$ at the interface 
\begin{equation}
\langle k_{z}^{2}\rangle=\frac{\left(12\right)^{2/3}}{4}\left(\frac{eEm}{\hbar^{2}}\right)^{2/3},
\end{equation}
where the field at the electron concentration $N$ can be estimated using the electrical neutrality of the total interface as: 
\begin{equation}
E=\frac{2\pi eN}{\epsilon }.
\end{equation}
For this choice of the wavefunction shape we have with a high accuracy 
\begin{equation}
\langle z^{2}\rangle^{1/2}\approx 0.7d_{\rm wf},
\end{equation}
where $d_{\rm wf}$ is the full width of the $\psi(z)-$function  
at half-maximum. Therefore, $\langle k_{z}^{2}\rangle\approx 6/d_{\rm wf}^{2}.$
In our previous studies using optical probing of the interface we reported the values of $d_{\rm wf}$~\cite{WaveFunctionZnO}. 
Note that the above formulas yield at $N=10^{12}\mbox{ cm}^{-2}$ the width $\langle z^{2}\rangle^{1/2}\approx 4.0$ nm, 
close to Ref.~\cite{WaveFunctionZnO}.
\begin{table}[!thb]
\caption{\label{tab:tableB} Tabulated FWHM of the wavefunction and the corresponding $\xi$-parameter.}
\begin{ruledtabular}
\begin{tabular}{ccc}
 $N$ & FWHM      & $\xi$\\
 $[10^{11}\text{ cm}^{-2}]$   &[nm]      &      [nm$^{-1}$]\\ 
	\hline
 1.3 & 8.2            & 0.60 \\
 2.04 & 7.2            & 0.68 \\
 3.09 & 6.2            & 0.79 \\
6.09 & 4.8            & 1.02 \\
7.25 & 4.6            & 1.06 \\
\end{tabular}
\end{ruledtabular}
\end{table}

Filled black squares in Fig. \ref{SI1} represent FWHM $d_{\rm wf}$ as a function of electron density $N$. 
The phenomenological fit  for $N$ expressed in the units of $10^{11}$ cm$^{-2},$ $d_{\rm wf}=4.16+6.62\exp(-N/2.67)$  (green line) describes well the $d_{\rm wf}$ 
dependence on $N$. Filled circles represent the samples used in the current work. Since the heterostructure design 
of the current samples is the same as of those used in Ref.~\cite{WaveFunctionZnO}, the $N-$dependence of the 
wavefunction width is estimated using the same phenomenological dependence. 
Knowing the FWHM of the electron states in our samples we find the corresponding $\xi$-parameter, summarized in the Table \ref{tab:tableB}.


\subsection{\bf 7. Table of various materials}
Summary of electron system parameters for Fig.4c of the main text. 
\begin{table}[!thb]
\caption{\label{tab:table1} Comparison of different semiconductors. $k_{\rm SOC}= m\alpha/\hbar^2$}
\begin{ruledtabular}
\begin{tabular}{ccccc}
 material system      & SOC coefficient               & $n$                                   &      $m/m_{0}$           & $k_{\rm SOC}$\\
                                  & [meV$\cdot{\rm \AA}$]      &$[10^{11}\text{ cm}^{-2}]$       &                                 &  [$\mu$m$^{-1}$]\\ 
	\hline
 InSb/InAlSb~\cite{SOIInSb}            & 130                                         & 3                                        &  0.014               &2.4\\
 InGaAs/InAlAs~\cite{SOIInGaAs}        & 70                                           & 20                                     & 0.046              &4.25\\
 InAs/AlSb~\cite{SOIInAs}             & 60                                           & 15                                     & 0.04              &3.17\\
GaN/AlGaN~\cite{SOIGaN}            &   6                                           & 10                                     & 0.22                &1.74\\
 GaAs (electrons)~\cite{SOIGaAs}   &  1-4                                         & 5                                       & 0.067              &0.35\\
 SiGe/Si/SiGe~\cite{SOISiGe}        & 0.05                                        & 5                                       & 0.19                &0.0125\\
\end{tabular}
\end{ruledtabular}
\end{table}

\end{document}